\documentclass[
    aps,
    prl,
    a4paper,
    10pt,
    reprint,
    twocolumn,
    nofootinbib,
    longbibliography,
    amsmath,amssymb,amssymb,
    floatfix,
    superscriptaddress,
    nofootinbib,
    noeprint
]{revtex4-2}

\usepackage[english]{babel}
\usepackage[utf8]{inputenc}
\usepackage[T1]{fontenc}
\usepackage{graphicx}
\usepackage{dcolumn}
\usepackage{bbm}
\usepackage{bm}
\usepackage{braket}
\usepackage{comment}
\usepackage[dvipsnames]{xcolor}
\usepackage[colorlinks,%
            linkcolor=BrickRed,%
            citecolor=MidnightBlue,%
            urlcolor=MidnightBlue]{hyperref}
\usepackage[normalem]{ulem}

\newcounter{ga}

\newcounter{ls}

\newcounter{pr}

\begin{document}

\title{Universality, Robustness, and Limits of the Eigenstate Thermalization Hypothesis in Open Quantum Systems}%

\author{Gabriel Almeida}
\email{gabriel.m.almeida@tecnico.ulisboa.pt}
\affiliation{CeFEMA-LaPMET, Departamento de Física, Instituto Superior Técnico, Universidade de Lisboa, Av. Rovisco Pais, 1049-001 Lisboa, Portugal}

\author{Pedro Ribeiro}
\email{ribeiro.pedro@tecnico.ulisboa.pt}
\affiliation{CeFEMA-LaPMET, Departamento de Física, Instituto Superior Técnico, Universidade de Lisboa, Av. Rovisco Pais, 1049-001 Lisboa, Portugal}
\affiliation{Beijing Computational Science Research Center, Beijing 100193, China}

\author{Masudul Haque}
\affiliation{Institut f\"ur Theoretische Physik, Technische Universit\"at Dresden, 01062 Dresden, Germany}

\author{Lucas  S\'a}
\email{ld710@cam.ac.uk}
\affiliation{TCM Group, Cavendish Laboratory, University of Cambridge, JJ Thomson Avenue, Cambridge CB3 0HE, UK\looseness=-1}

\begin{abstract}
The eigenstate thermalization hypothesis (ETH) underpins much of our modern understanding of the thermalization of closed quantum many-body systems. Here, we investigate the statistical properties of observables in the eigenbasis of the Lindbladian operator of a Markovian open quantum system. We demonstrate the validity of a \emph{Lindbladian ETH ansatz} through extensive numerical simulations of several physical models. To highlight the robustness of Lindbladian ETH, we consider what we dub the dilute-click regime of the model, in which one postselects only quantum trajectories with a finite fraction of quantum jumps. The average dynamics are generated by a non-trace-preserving Liouvillian, and we show that the Lindbladian ETH ansatz still holds in this case. On the other hand, the no-click limit is a singular point at which the Lindbladian reduces to a doubled non-Hermitian Hamiltonian and Lindbladian ETH breaks down.
\end{abstract}

\maketitle

\emph{Introduction.}---%
Great insights into chaotic closed many-body quantum systems arise when we model their complex interactions as random. This is because, in many ways, the Hamiltonian of these systems exhibits the same statistical behavior as a large random matrix. Particularly noteworthy is the modern reinterpretation of the eigenstate thermalization hypothesis (ETH)~\cite{deutsch1991quantum,srednicki1994chaos}:
The random matrix theory (RMT) structure of eigenstates can be upgraded to explain thermalization in closed quantum systems.

The ETH states that the matrix elements of a local observable $O$ in the energy eigenbasis $\ket{n}$, defined by $H\ket{n}=E_n \ket{n}$, follow the ansatz
\begin{equation}
\braket{m|O|n} =\overline{O}(\bar{E}) \delta_{mn} + S_{mn} \sqrt{\frac{\overline{O^2}(\bar{E},\omega)}{\mathcal{D}}},
\label{eq:closed-eth}
\end{equation}
where $\overline{O}$ (resp. $\overline{O^2}$) is the microcanonical averaged value of $O$ (resp. $O^2$), $\bar{E} = (E_m + E_n)/2$, $\omega = E_m - E_n$, $\mathcal{D}$ is the Hilbert space dimension and $S_{mn}$ is a random variable with zero mean and unit variance. ETH ensures that fluctuations around the microcanonical value of the observable are suppressed in the thermodynamic limit, thus matching the temporal average of $O$ with the value expected from statistical mechanics. ETH is thus central to understanding thermalization, as it provides a mechanism by which isolated quantum systems can thermalize and there is extensive numerical evidence supporting it~\cite{rigol2008thermalization, rigol2010quantum,Polkovnikov_RMP2011,santos2013,Beugeling_scaling_PRE14,nandkishore2015,alessio2016,mondaini2016eigenstate}. In contrast, violations of ETH can occur in specific contexts, such as integrable and many-body localized systems \cite{altshuler1997quasiparticle,basko2006metal,gornyi2005interacting,Ziraldo_Santoro_relaxation_PRB2013, Ikeda_Ueda_PRE2013_LiebLiniger, Beugeling_scaling_PRE14, Alba_PRB2015, li2015many, ArnabSenArnabDas_PRB16, Magan_randomfreefermions_PRL2016, HaqueMcClarty_SYKETH_PRB2019, Mierzejewski_Vidmar_PRL2020,noh2021eigenstate}, where an extensive number of conservation laws prevent thermalization; systems exhibiting quantum many-body scars \cite{bernien2017Nat,turner2018NatPhys,shiraishi2017,serbyn2021,bernevig2022}, in which a subset of quantum states avoid ergodicity; and systems showing Hilbert space fragmentation \cite{khemani2020localization,sala2020ergodicity,moudgalya2022hilbert,khudorozhkov2022hilbert,adler2024observation}.

Given the profound consequences of ETH on the dynamics of closed quantum systems, it is natural to ask if there is an analogous result for open quantum systems, which interact with their environment and give rise to mixed states. In recent years, some progress has been made in this direction~\cite{arovas-sondhi2019,hamazaki2022lindbladian,cipolloni2024non,non-hermitian-eth,richter2024,garcia2025scars}.   
In some limits, open quantum systems can be modeled by non-Hermitian Hamiltonians \cite{ashida2020non}. While not conserving probability, such a description can be appropriate for short-time evolutions or when postselecting jump-free quantum trajectories. ETH still holds for these systems~\cite{non-hermitian-eth}, provided that one computes the matrix elements with the basis of right (or left) eigenvectors, $\ket{r_n}$ (or $\ket{l_n}$), and one introduces a term that accounts for the nonorthogonality of the eigenvectors. The non-Hermitian ETH ansatz is then~\cite{non-hermitian-eth}
\begin{equation}
    \braket{r_m|O|r_n} = \overline{O} \braket{r_m|r_n} + S_{mn} \sqrt{\frac{\overline{O^2}}{\mathcal{D}}},
    \label{eq:non-hermitian-eth}
\end{equation}
where the $\overline{O}$ and $\overline{O^2}$ dependence on the eigenvalues is implicit. If the Hamiltonian is Hermitian, we have $\braket{r_m|r_n}=\delta_{mn}$ and recover Eq.~\eqref{eq:closed-eth}. This choice of matrix elements is crucial since it was proven~\cite{cipolloni2024non} that there is no ETH for biorthogonal matrix elements such as $\braket{l_m|O|r_n}$.

A more complete description of open quantum systems employs a quantum master equation for the system's reduced density matrix $\rho$, $\dot{\rho}=\mathcal{L}[\rho]$, where $\mathcal{L}$ is the Liouvillian superoperator~\cite{breuer2002theory}. If the environment is Markovian, the Liouvillian takes the Lindblad form \cite{lindblad1976generators,gorini1976completely}:
\begin{equation}
    \mathcal{L}[\rho]=-i[H,\rho]+ \sum_{k=1}^r \left( W_k \rho W_k^\dagger - \frac{1}{2} \left\{W_k^\dagger W_k, \rho\right\}\right).
    \label{eq:lindbladian}
\end{equation}
Here, $H$ is the Hamiltonian of the system and $W_k$ (with $k=1,\dots, r$) are the jump operators describing the coupling to the environment. 

Whereas in equilibrium the eigenstates of the Hamiltonian and the (thermal) steady state coincide, nonequilibrium dynamics offer the more intriguing possibility of independently probing the structure of the eigenstates of the dynamical generator and steady state. The latter allows a straightforward extension~\cite{arovas-sondhi2019} of ETH, even far from equilibrium~\cite{richter2024}, by writing it as the exponential of an effective Hamiltonian, to which Eq.~(\ref{eq:closed-eth}) applies. In this Letter, we instead focus on the former, with our main results summarized as follows. Inspection of the time evolution of observables suggests that the role of matrix elements in the Hamiltonian eigenbasis is played in this context by overlaps of observables with (operator) eigenstates of the Lindbladian superoperator. Treating these eigenstates as random vectors leads to a \textit{Lindbladian ETH ansatz}, whose validity and universality we test in three increasingly realistic physical models. Moreover, we conjecture that the ansatz is robust to any deformations that preserve the two-copy structure of the Lindbladian and show this explicitly for non-trace-preserving dynamics. Such dynamics can be obtained by averaging over quantum trajectories postselected to have a fixed rate of quantum jumps---what we dub the \emph{dilute-click limit} and which continuously interpolates between Lindbladian dynamics and the no-click limit \cite{liu2025lindbladian, gupta2024quantum}. In the latter, the Liouvillian decouples into two non-Hermitian Hamiltonians, and we show that Lindbladian ETH breaks down. In this limit, the eigenstates have additional structure, which adds a correction to the ansatz~\cite{non-hermitian-eth}.

\emph{Lindbladian ETH ansatz.}---%
In the Lindbladian setting, the time evolution of an observable $O$ is given by
\begin{equation}
    \braket{O}_t = \operatorname{Tr}[O e^{\mathcal{L} t}[ \rho]]= \sum_\mu e^{\lambda_\mu t}\operatorname{Tr}[O R_\mu] \operatorname{Tr}[L_\mu^\dagger \rho],
    \label{eq:dynamics}
\end{equation}
where $\lambda_\mu$ are the eigenvalues of $\mathcal{L}$, while $R_\mu$ and $L_\mu$ are the respective right and left eigenoperators: $\mathcal{L}[R_\mu] = \lambda_\mu R_\mu$ and $\mathcal{L}^\dagger[L_\mu] = \lambda_\mu^\ast L_\mu$. Additionally, we choose the normalization $\operatorname{Tr}[L_\mu^\dagger R_\nu]=\delta_{\mu \nu}$ and $\|R_\mu\|^2 = \operatorname{Tr}[R_\mu^\dagger R_\mu]=1$. To separate the scaling of $O$ and $\rho$ and to work with unit vectors, we can rewrite Eq.~\eqref{eq:dynamics} as
\begin{equation}
\label{eq:overlaps_normalized}
    \frac{\braket{O}_t}{\|O\| \cdot\|\rho\|} = \sum_\mu e^{\lambda_\mu t} \frac{\operatorname{Tr}[O R_\mu]}{\|O\|} \frac{\operatorname{Tr}[L_\mu^\dagger \rho]}{\|L_\mu\|\cdot \|\rho\|} \|L_\mu\|.
\end{equation}
Therefore, we identify three important quantities for the dynamics: $\operatorname{OR}_\mu=\frac{\operatorname{Tr}[O R_\mu]}{\|O\|}$, $\operatorname{L\rho}_\mu = \frac{\operatorname{Tr}[L_\mu^\dagger \rho]}{\|L_\mu\|\cdot \|\rho\|}$ and $\operatorname{LL}_\mu = \|L_\mu\|^2= \operatorname{Tr}[L_\mu^\dagger L_\mu]$. Related quantities have been studied in Ref.~\cite{hamazaki2022lindbladian} that considered a disordered Ising chain with dissipation. For small disorder, the authors found that $\operatorname{OR}$ and $\operatorname{L \rho}$ are random variables with a complex Gaussian distribution, while $\operatorname{LL}$ is a random variable with a $1/\gamma_2$ distribution.
In this work, we show that these distributions hold for generic Markovian open quantum systems and, based on them, propose a Linbladian ETH ansatz. 
For a generic chaotic Lindbladian, we thus have the following ansatz for the overlaps appearing in Eq.~(\ref{eq:overlaps_normalized}) (see also Ref.~\cite{hamazaki2022lindbladian}):
\begin{align}
    \label{eq:ansatz_OR}
    \operatorname{OR}_\mu &= s_\mu \frac{f_1(\lambda_\mu)}{\mathcal{D}},\\
    \label{eq:ansatz_Lrho}
    \operatorname{L\rho}_\mu&= s_\mu' \frac{f_2(\lambda_\mu)}{\mathcal{D}},\\
    \label{eq:ansatz_LL}
    \operatorname{LL}_\mu &= r_\mu \mathcal{D}^2 f_3(\lambda_\mu).
\end{align}
Here, $\mathcal{D}$ is the many-body Hilbert space dimension, $s_\mu$ and $s_\mu'$ are random variables with a complex normal distribution with mean zero and unit variance, $r_\mu$ is a random variable with a $1/\gamma_2$ distribution, and $f_i(\lambda_\mu)$ are smooth functions of the Liouvillian eigenvalues~\cite{hamazaki2022lindbladian}. This ansatz is justified by treating the $\mathcal{D}^2$-dimensional eigenvectors of $\mathcal{L}$ as complex Gaussian random vectors. For example, for the $\operatorname{OR}$ overlap we have:
\begin{align}
    \overline{\operatorname{Tr}[OR_n]} &= \sum_{i} O_i^\ast\overline{R_{n,i}} = 0,
    \\
    \overline{|\operatorname{Tr}[OR_n]|^2} &= \sum_{i,j} O_i^\ast O_j  \overline{R_{n,i}^\ast R_{n,j}}
    \nonumber \\
    &= \sum_{ij} O_i^\ast O_j \frac{\delta_{ij}}{\mathcal{D}^2} = \frac{\|O\|^2}{\mathcal{D}^2}.
\end{align}
With the normalization $\|R_\mu\|^2=1$, $\operatorname{LL}_\mu$ is the diagonal Chalker-Mehlig overlap of eigenvectors~\cite{mehlig1998PRL,mehlig2000JMP}, whose inverse was proven in Ref.~\cite{bourgade2020} to follow a $\gamma_2$ distribution.
Below, we will verify the applicability of Eqs.~(\ref{eq:ansatz_OR})--(\ref{eq:ansatz_LL}) by computing the local scalings and distributions of $\operatorname{OR}$, $\operatorname{L \rho}$, and $\operatorname{LL}$ in different physical models. 

\begin{figure}[tbp]
    \centering
    \includegraphics[width=\columnwidth]{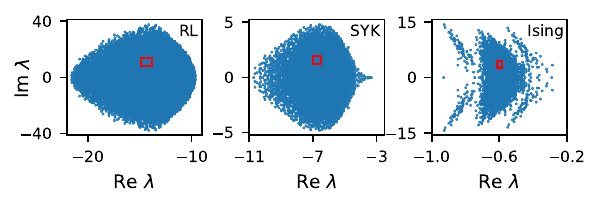}
    \caption{Spectrum of a random Liouvillian (left), the dissipative SYK model (center), and the damped Ising model (right). We study the local scalings and distributions of $\operatorname{OR}$, $\operatorname{L\rho}$, and $\operatorname{LL}$ in a small region of the spectrum (red box) defined by $[\mu_x - w_x/5, \mu_x + w_x/5]$ $\times$ $[2w_y/5, 4w_y/5]$, where $\mu_x$ is the average value of $\operatorname{Re}\lambda$ and $w_x$ ($w_y$) is the standard deviation of $\operatorname{Re}\lambda$ ($\operatorname{Im}\lambda$).}
    \label{fig:spec-region}
\end{figure} 

\emph{Physical models.}---%
To test the universality of the Lindbladian ETH ansatz, we study three disordered physical models with different degrees of locality, namely, a random-matrix Liouvillian~\cite{denisov2019prl,can2019,can2019b,sa2020spectral,denisov2021}, the dissipative Sachdev-Ye-Kitaev (SYK) model~\cite{sa2022PRR,Kulkarni2022,garcia2022,kawabata2022,garcia2025PRB}, and a damped disordered Ising chain~\cite{hamazaki2022lindbladian}.
The Liouvillian spectra of the three models are shown in Fig.~\ref{fig:spec-region}.

We begin with the completely disordered case of random Liouvillians~\cite{sa2020spectral}. To obtain a random Liouvillian, we draw a $\mathcal{D}\times\mathcal{D}$ random Hermitian matrix $H$ in Eq.~(\ref{eq:lindbladian}) from the Gaussian unitary ensemble (GUE), $P_\mathcal{D}(H) \propto \exp\left\{-\frac{1}{2}\operatorname{Tr}[H^2]\right\}$,
and $r$ random non-Hermitian matrices $W_k$ from the Ginibre unitary ensemble (GinUE), $P_\mathcal{D}(W_k) \propto \exp\left\{-\frac{g^2}{2}\operatorname{Tr}[W_k^\dagger W_k]\right\}$,
from which we remove the trace, $W_k \to W_k - \operatorname{Tr}[W_k]$. The variance $g \equiv g_\text{eff} (4 r \mathcal{D})^{-1/4}$ measures the dissipation strength~\cite{sa2020spectral}. In the following, we set $r=5$ and $g_\text{eff} =0.8$.

Second, we consider the dissipative SYK model of $N\in2\mathbb{Z}$ Majorana fermions, which satisfy the Clifford algebra $\{\chi_i, \chi_j\}=\delta_{ij}$, with infinite-range four-body interactions in zero dimensions. Following Ref.~\cite{sa2022PRR}, the Hamiltonian of this model is~\cite{kitaev2015,maldacena2016}
\begin{equation}
    H=\sum_{i<j<k<l} J_{ijkl} \chi_i \chi_j \chi_k \chi_l
\end{equation}
and the $M=m N$ jump operators are given by
\begin{equation}
    W_m = i \sum_{i<j} \ell_{m,ij} \chi_i \chi_j.
\end{equation}
The couplings $J_{ijkl}$ and $\ell_{m,ij}$ are independent Gaussian random variables with zero mean and variance $\braket{J^2_{ijkl}} = \frac{3! J^2}{N^3}$ and $\braket{\lvert \ell_{m,ij} \rvert^2} = \frac{\gamma^2}{N^2}$
respectively. The coefficients $J_{ijkl}$ must be real to ensure the Hermiticity of the Hamiltonian, while $\ell_{m,ij}$ are taken to be complex. We set  $J=m=\gamma=1$.

Finally, we employ the damped tilted-field Ising model of Ref.~\cite{hamazaki2022lindbladian}, which has local interactions and dissipation but spatial disorder. The Hamiltonian is given by
\begin{equation}
    H=\sum_{i=1}^{L-1} \sigma_i^z \sigma_{i+1}^z + g\sum_{i=1}^L  \sigma_i^x +\sum_{i=1}^L h_i \sigma_i^z,
\end{equation}
where $\sigma_i^\alpha$ ($\alpha=x,y,z$) are the standard Pauli matrices on site $i$, $g=0.9$ is a constant transverse field, and the disordered longitudinal field $h_i$ is taken randomly from $[-h,+h]$ with $h=0.2$; the jump operators are $W_k=\sqrt{\mu/2} \sigma_k^-$, with $k=1,2,\dots, L$ and $\mu=0.4$.  We use open boundary conditions.

\begin{figure}[t]
    \centering
    \includegraphics[width=0.9\columnwidth]{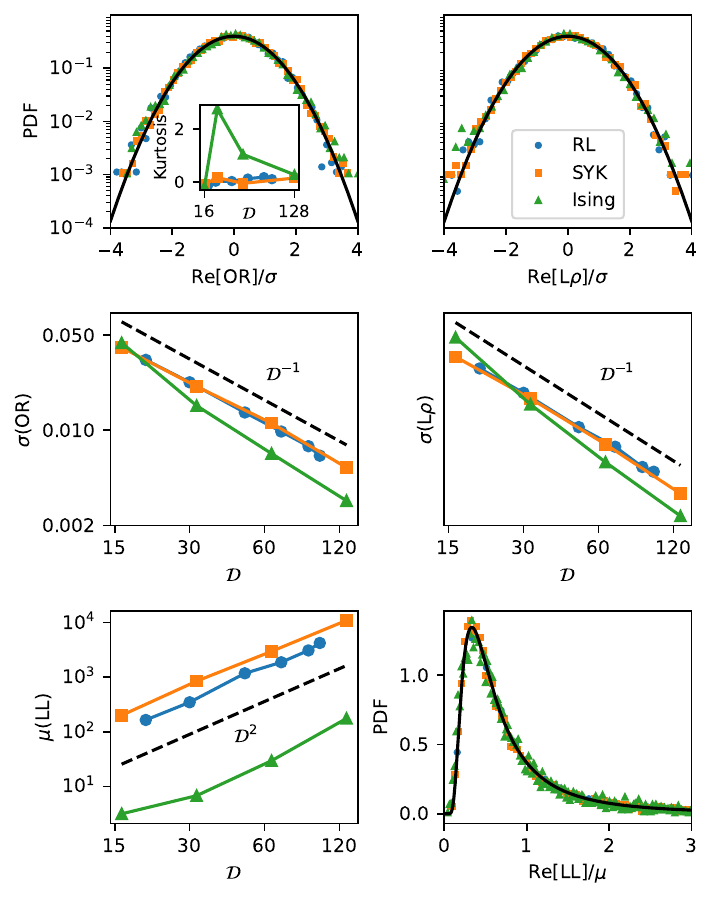}
    \caption{Verification of the Lindbladian ETH ansatz for random Liouvillians, the dissipative SYK model (with observable $O=\chi_1\chi_2\chi_3\chi_4$), and a damped Ising chain.
    The Hilbert space dimension is $\mathcal{D}=100$ for the random Liouvillian and $\mathcal{D}=128$ for the SYK and Ising models.
    (Top row) Distribution of $\operatorname{OR}$ (left) and $\operatorname{L\rho}$ (right), normalized by the respective standard deviations. The colored dots are histograms for the three different models, which are compared with a standard Gaussian distribution (black line). The inset shows the excess kurtosis, which is zero for a Gaussian distribution, as a function of the Hilbert space dimension; for all models, it is zero or decreases with system size.
    (Middle row) Scaling of the width of the distribution with the Hilbert space dimension $\mathcal{D}$. In all three models, $\operatorname{OR}$ (left) and $\operatorname{L\rho}$ (right) scale as $\mathcal{D}^{-1}$ (dashed line). 
    (Bottom row, left) Scaling of the mean of $\operatorname{LL}$, compared with the RMT prediction, $\mathcal{D}^2$ (dashed line). 
    (Bottom row, right) Distribution of $\operatorname{LL}$ normalized by its mean. The black line is the $1/\gamma_2$ distribution, which is followed in all three physical models.
    }
    \label{fig:eth}
\end{figure}

\begin{figure}[t]
    \centering
    \includegraphics[width=0.95\linewidth]{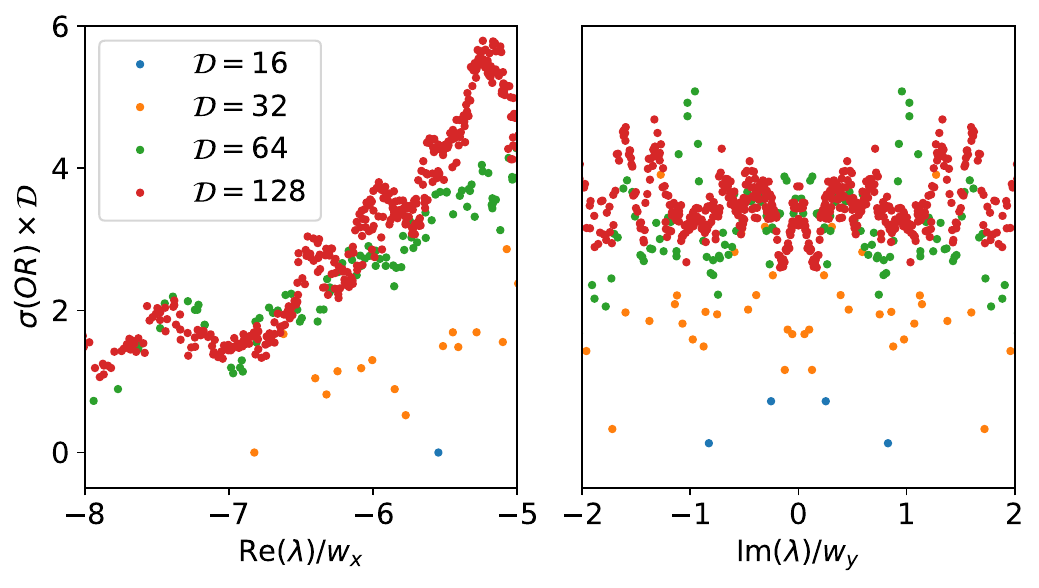}
    \caption{Form function $f_1$ [defined in Eq.~\eqref{eq:ansatz_OR}] for a single realization of the SYK model. We compute the local standard deviation of $\operatorname{Re}[\operatorname{OR}]$ rescaled by $\mathcal{D}$. The left panel shows a cut parallel to the real axis, taken at $\operatorname{Im} \lambda \in [0.45, 0.55]w_y$, while the right panel shows a cut parallel to the imaginary axis at $\operatorname{Re}\lambda \in \mu_x + [- 0.05, 0.05] w_x$. In both cases, up to the numerically available system sizes, the data is consistent with a collapse onto a smooth function as the system size increases.}
    \label{fig:f1-SYK}
\end{figure}

\emph{Universality of Lindbladian ETH.---}%
We now show the universality of Lindbladian ETH by computing the local scalings and distributions of $\operatorname{OR}$, $\operatorname{L\rho}$, and $\operatorname{LL}$ in the three models introduced above. 
For the random Liouvillian, the observable is $O=O_2 \otimes 1_{\mathcal{D}/2}$, where $O_2$ is a $2\times2$ random matrix from the Gaussian orthogonal ensemble and the initial state is $\rho=\ket{0}\bra{0}$ for an arbitrary computational basis state $\ket{0}$ in the many-body Hilbert space. In the SYK model, we consider two different observables, $O=i\chi_1 \chi_2$ and $O=\chi_1 \chi_2 \chi_3 \chi_4$, and the initial state is $\rho=\ket{\Omega}\bra{\Omega}$ where $\ket{\Omega}$ is the fermionic vacuum of a single copy of the SYK model. In the Ising model, the observable is $O=\sigma^x_{L/2}$ and the initial state is $\rho=\ket{\uparrow \dots \uparrow}\bra{\uparrow\dots \uparrow}$, where $\ket{\uparrow \dots \uparrow}$ is the $L$-fold tensor product of the $+1$ eigenstate of $\sigma^z$.

We select a small region in the bulk of the complex spectrum (see Fig.~\ref{fig:spec-region}), compute the three quantities for approximately $10^5/\mathcal{D}^2$ independent realizations, and determine their average, variance, and full distribution. We reach $\mathcal{D}=100$ for random Liouvillians and $\mathcal{D}=128$ for the SYK and Ising models, respectively, and show the distributions for this largest size. The results are shown in Fig.~\ref{fig:eth} and confirm the validity of the Lindbladian ETH ansatz: in all three models, $\operatorname{OR}$ and $\operatorname{L\rho}$ have a complex Gaussian distribution and scale as $\mathcal{D}^{-1}$, while $\operatorname{LL}$ has a $1/\gamma_2$ distribution and scales as $\mathcal{D}^2$. Moreover, the deviation for the expected distributions decreases with system size, as shown for OR in the inset of the upper-left panel of Fig.~\ref{fig:eth}.

We also assess the smoothness of the functions $f_i$ appearing in the Lindbladian ETH ansatz. Specifically, we compute the local standard deviation over an elliptical neighborhood (with semi-axis $0.2 w_x$ and $0.2 w_y$) for each eigenvalue. This analysis is performed for a single realization of the SYK model. The results, shown in Fig.~\ref{fig:f1-SYK}, indicate the convergence to a smooth function as the system size increases.

\emph{Robustness of Lindbladian ETH.}---%
We now show that Lindbladian ETH is robust and is still valid when trace preservation is relaxed. To this end, we introduce a deformed Liouvillian given by
\begin{equation}
    \label{eq:L_alpha}
    \mathcal{L}_\alpha[\rho] = -i K \rho + i \rho K^\dagger + \alpha \sum_k W_k \rho W_k^\dagger,
\end{equation}
where $K=H - \frac{i}{2}\sum_k W_k^\dagger W_k$ is an effective non-Hermitian Hamiltonian. Here,  $\alpha\in[0,1]$ quantifies the trace preservation (i.e., probability conservation) of the dynamics, with $\mathcal{L}_{\alpha=1}$ the usual Lindbladian. The regime $\alpha<1$ corresponds to postselecting only trajectories with a fraction $\alpha$ of jumps (``dilute-click regime'') \cite{liu2025lindbladian, gupta2024quantum}, as shown in the End Matter. The dynamics under $\mathcal{L}_{\alpha = 0}$ arises in the no-click limit~\cite{dalibard1992wave,zoller1987quantum,visser1995solution}, where only quantum trajectories without quantum jumps are postselected. In this limit, $\mathcal{L}_{0}$ reduces to two decoupled replicas of a non-Hermitian Hamiltonian. 

\begin{figure}[t]
    \centering
    \includegraphics[width=\columnwidth]{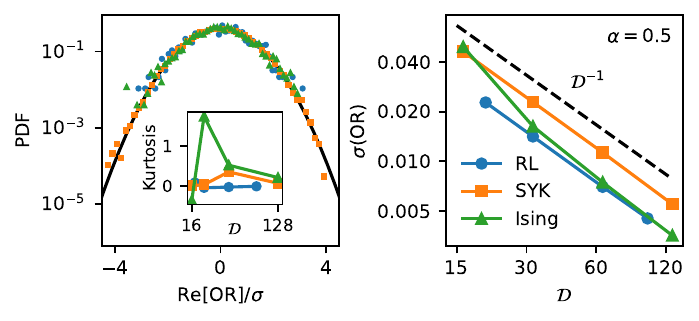}
    \caption{Distribution (left) and scaling of the variance with Hilbert space dimension (right) of $\operatorname{Re[OR]}$ for $\alpha=0.5$ in the three physical models. Here, we use $\mu=0.8$ for the Ising model to converge faster to the RMT regime. Despite the loss of trace preservation, the scaling of the variances is still that predicted by the Lindblad ETH ansatz for all models. Moreover, the full distribution follows a Gaussian, while there are deviations in the tails for the Ising model. In the inset of the left panel, we compute the excess kurtosis, which is zero for a Gaussian distribution, as a function of the Hilbert space dimension; for all models, it decreases with system size.
    }
    \label{fig:alpha-05}
\end{figure}

The distribution and scaling of $\operatorname{OR}$ for $\alpha=0.5$ are shown in Fig.~\ref{fig:alpha-05}. 
For random Liouvillians and the SYK model, $\operatorname{OR}$ is still Gaussian distributed and scales with $\mathcal{D}^{-1}$, as predicted by our Lindbladian ETH ansatz.
For the damped Ising model, the scaling matches the prediction but there are deviations in the tails of the distribution. To investigate this discrepancy, we compute the excess kurtosis of the distribution, which is zero for a Gaussian, as a function of $\mathcal{D}$ for all three models, see the inset of Fig.~\ref{fig:alpha-05}. For the random Liouvillian, it is consistent with zero for all system sizes, while for the SYK model it decreases to close to zero at $\mathcal{D}\sim100$. For the Ising model, the available system sizes do not allow us to reach a definitive conclusion, but the data is consistent with an approach to Gaussianity in the thermodynamic limit, albeit at a much slower rate than for the other two models (which are more disordered and less constrained by locality).

A similar agreement is found for $\operatorname{L\rho}$ and $\operatorname{LL}$ and for other values of $\alpha>0$. More generally, we expect Lindbladian ETH to hold as long as the dynamical generator couples the two copies of the Hilbert space (i.e., the bra and the ket of the density matrix), as a generic Hermiticity-preserving generator does.

\begin{figure}[t]
    \centering
    \includegraphics[width=\linewidth]{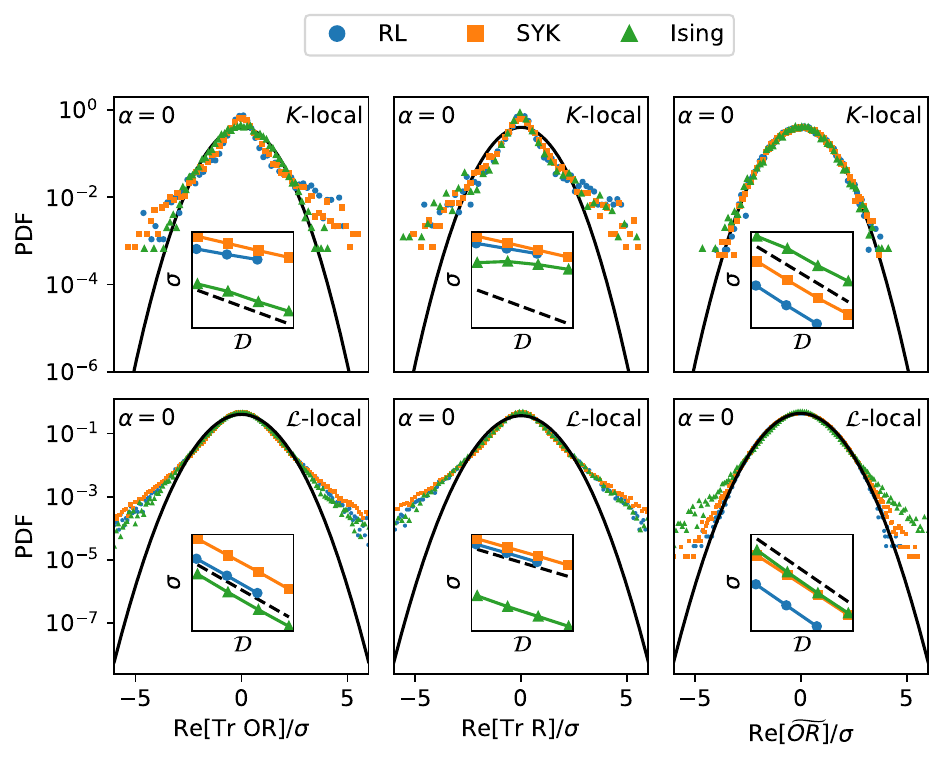}
    \caption{Distribution of $\operatorname{Tr}[OR]$, $\operatorname{Tr}R$, and $\widetilde{\operatorname{OR}}\equiv \operatorname{Tr}[OR] - \overline{O} \operatorname{Tr} R$ for $\alpha=0$ in random Liouvillians, the dissipative SYK model and the damped Ising chain. In the top row, the distributions are computed in a small region of the spectrum of $K$, while in the bottom they are computed locally in the spectrum of $\mathcal{L}_0$. (Insets) The scaling of the variance of the distributions with the Hilbert space dimension is compared with $\mathcal{D}^{-1/2}$ (black dashed line). Here, the eigenvalue window is chosen as $[\mu_x - w_x/5, \mu_x + w_x/5] \times [w_y/5, 2 w_y/5]$ and for the SYK model we take $O=i\chi_1 \chi_2$ as the observable.}
    \label{fig:alpha-0}
\end{figure}

\emph{Limits of Lindbladian ETH.}---%
The condition mentioned in the previous paragraph fails for $\alpha=0$, where the jump contribution vanishes, the two copies decouple, and the Liouvillian corresponds to a doubled non-Hermitian Hamiltonian. The right eigenoperators of $\mathcal{L}_0$ are given by $R_{mn}=\ket{r_m}\bra{r_n}$ and have eigenvalues $\lambda_{mn}=-i (\varepsilon_m - \varepsilon^*_n)$, where $\ket{r_n}$ is the right eigenvector of $K$ with eigenvalue $\varepsilon_n$, $K|r_n\rangle=\varepsilon_n|r_n\rangle$. Due to this additional structure, and using Eq.~\eqref{eq:non-hermitian-eth}~\cite{non-hermitian-eth}, the right-eigenvector overlap $\operatorname{Tr}[O R_{mn}]$ becomes
\begin{align}
    \operatorname{Tr}[O R_{mn}] 
    &= \overline{O} \operatorname{Tr}[R_{mn}] + S_{mn} \sqrt{\frac{\overline{O^2}}{\mathcal{D}}}.
    \label{eq:eth-alpha0}
\end{align}
When $\varepsilon_n \approx \varepsilon_m$ (i.e., locally within the spectrum of $K$), this expression is a decomposition of $\operatorname{Tr}[OR]$ into a non-Gaussian fluctuation with heavy tails, $\operatorname{Tr}R$, and a Gaussian fluctuation, $\widetilde{OR}\equiv \operatorname{Tr}[OR] - \overline{O} \operatorname{Tr} R$, that scales as $\mathcal{D}^{-1/2}$. This behavior agrees with the non-Hermitian ETH prediction \cite{non-hermitian-eth} and is confirmed for all models in the top row of Fig.~\ref{fig:alpha-0}.

In contrast, when $\varepsilon_n$ and $\varepsilon_m$ are far apart---which is typically the case in a small window around $\lambda_{nm}$ (i.e., locally within the spectrum of $\mathcal{L}_0$)---even $\widetilde{OR}$ becomes non-Gaussian; see the bottom row of Fig.~\ref{fig:alpha-0}. Consequently, the $\operatorname{OR}$ overlap becomes generically non-Gaussian and the Lindbladian ETH ansatz breaks down at $\alpha=0$ (in the End Matter we provide further evidence that the ansatz breaks down only in the no-click limit). Note, however, that even though the distribution is non-Gaussian, its variance still has the predicted scaling $\mathcal{D}^{-1/2}$.

\emph{Conclusions.---}%
In this work, we studied the quantities that govern the dissipative dynamics of generic observables across several physical models. Our results show that these quantities exhibit universal scalings and distributions, similarly to ETH in isolated systems. By introducing a deformation parameter $\alpha \in [0,1]$, we found that the ansatz is robust against the loss of trace preservation ($0 < \alpha < 1$). However, $\alpha=0$ is a strikingly different case, as the Lindbladian decouples into two non-Hermitian Hamiltonians and Lindbladian ETH is destroyed.

This work paves the way for a deeper understanding of the structure of Lindblad eigenvectors in open quantum systems, paralleling the success of ETH in the context of closed systems.
Contrary to the case of isolated systems, in which ETH helps explain the relaxation of observables to a local thermal equilibrium, the fate of open quantum systems is always the steady state, which need not even be thermal. As such, the implications of Lindbladian ETH do not coincide with those of closed-system ETH.  Elucidating the structure of eigenstates is fundamental to our understanding of quantum formalisms, both for closed and for open systems.  For closed systems, the importance of ETH (and its extensions) extends well beyond its role in explaining the thermalization of local observables. 
Similarly, we expect the clarification of Lindbladian eigenstate structures to be essential groundwork for the future understanding of the physics of Markovian many-body open quantum systems.


\emph{Acknowledgments.---}%
GA and PR acknowledge support by FCT through Grant No. UID/CTM/04540 to the I\&D unit Centro de Física e Engenharia de Materiais Avançados. This work was produced with the support of INCD funded by FCT and FEDER under the project 01/SAICT/2016 nº 022153.  
This work was realized within the QuantERA ERA-NET project DQUANT: A Dissipative Quantum Chaos perspective on Near-Term Quantum Computing, supported by FCT-Portugal Grant Agreement No. 101017733.\footnote{\href{https://doi.org/10.54499/QuantERA/0003/2021}{https://doi.org/10.54499/QuantERA/0003/2021}} 
MH acknowledges support from the Deutsche Forschungsgemeinschaft under grant SFB 1143 (project-id 247310070). LS was supported by a Research Fellowship from the Royal Commission for the Exhibition of 1851. 

\bibliography{bibliography}

\section*{END MATTER}

\subsection{Dilute-click regime}
\label{appendix}

In this Appendix, we show that the deformed Liouvillian in Eq.~\eqref{eq:L_alpha} effectively reduces the jump rate by a factor of $\alpha$, meaning that quantum trajectories with jumps are accepted only with probability $\alpha$. This post-selection interpretation can be seen as follows.

For a small time step $\delta t$, the density matrix evolves as  
\begin{align}
    \rho(t+\delta t) \approx &(1-i K \delta t) \rho(t) (1 + i K^\dagger \delta t) \nonumber\\ 
    & + \alpha \delta t \sum_k W_k \rho(t) W_k^\dagger.
\end{align}  
The first term corresponds to evolution without jumps, while the second term accounts for the possibility of jumps occurring. To interpret this evolution probabilistically, we define the normalized (unit-trace) density matrices  
\begin{align}
    \rho^{(0)} &= \frac{(1-i K \delta t) \rho(t) (1 + i K^\dagger \delta t)}{p_0},\\
    \rho^{(k)} &= \frac{\alpha \delta t W_k \rho(t) W_k^\dagger}{\delta p_k},
\end{align}  
with the associated probabilities  
\begin{align}
    p_0 &= \operatorname{Tr} [(1-i K \delta t) \rho(t) (1 + i K^\dagger \delta t)],\\
    \delta p_k &= \alpha \delta t \operatorname{Tr}[W_k \rho(t) W_k^\dagger].
\end{align}  
Thus, the evolved density matrix takes the form
\begin{equation}
    \rho(t+\delta t) = p_0 \rho^{(0)} + \sum_k \delta p_k \rho^{(k)}.
\end{equation}  
This is a statistical mixture of $\rho^{(0)}$, which corresponds to the evolution if no jump occurs, and states $\rho^{(k)}$, which are the collapsed states following a jump associated with $W_k$. Unless $\alpha=1$, we have $p_0 + \sum_k p_k < 1$, meaning that the evolved density matrix has trace less than 1. The missing trace corresponds to the probability of trajectories with jumps that were discarded in the post-selection process. 

Note that, as in the no-click limit, the Liouvillian studied in the main text differs from the generator of the diluted click dynamics given above by a constant factor that is a non-linear function of the density matrix.  Therefore, although the evolution with the linear Liouvillian reproduces the correct density matrix, it is not able to capture the probability of the dilute click trajectories, similarly to the no-click limit.

Since $\delta p_k = \alpha\ \delta p_k|_{\alpha = 1}$, this evolution corresponds to, at each time step, accepting a jump only with probability $\alpha$. In the no-click limit, $\alpha = 0$, jumps are completely discarded and the system evolves under the non-Hermitian Hamiltonian $K$. For $\alpha = 1$, we recover the standard trace-preserving Lindblad evolution, where jumps occur with their original probability. For intermediate values $0 < \alpha < 1$, the system follows a ``diluted jump'' regime, where jumps occur less frequently than in the standard Lindblad evolution.

\subsection{No-click limit}
\label{appendix-alpha0}

\begin{figure}[t]
    \centering
    \includegraphics[width=0.65\columnwidth]{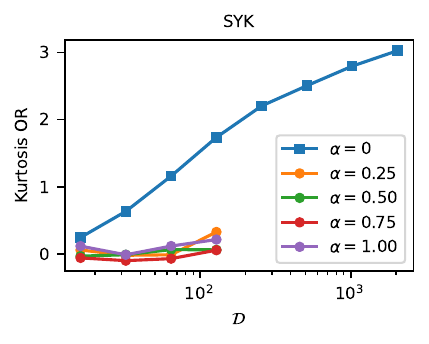}
    \caption{Excess kurtosis of the distribution of $\operatorname{OR}$ as function of the Hilbert space dimension in the dissipative SYK model (with observable $O=i \chi_1 \chi_2$) for different values of $\alpha$. A value of zero corresponds to a Gaussian distribution, which is attained for all $\alpha > 0$; for $\alpha=0$, the kurtosis increases rapidly with $\mathcal{D}$, which signals a strong deviation from Gaussianity and a breakdown of Lindbladian ETH.}
    \label{fig:kurtosis-alphas}
\end{figure}

In this Appendix, we further substantiate the claim that the distributions become non-Gaussian only at the singular point $\alpha=0$ (no-click limit), while Lindbladian ETH holds for any nonzero $\alpha$. In Fig.~\ref{fig:kurtosis-alphas}, we plot the excess kurtosis as a function of the Hilbert space dimension for the dissipative SYK model with different $\alpha$. It remains close to zero for $\alpha=0.25,\dots,1$, while it grows steeply with $\mathcal{D}$ at the singular point $\alpha=0$.

\begin{figure}[t]
    \centering
    \includegraphics[width=0.95\linewidth]{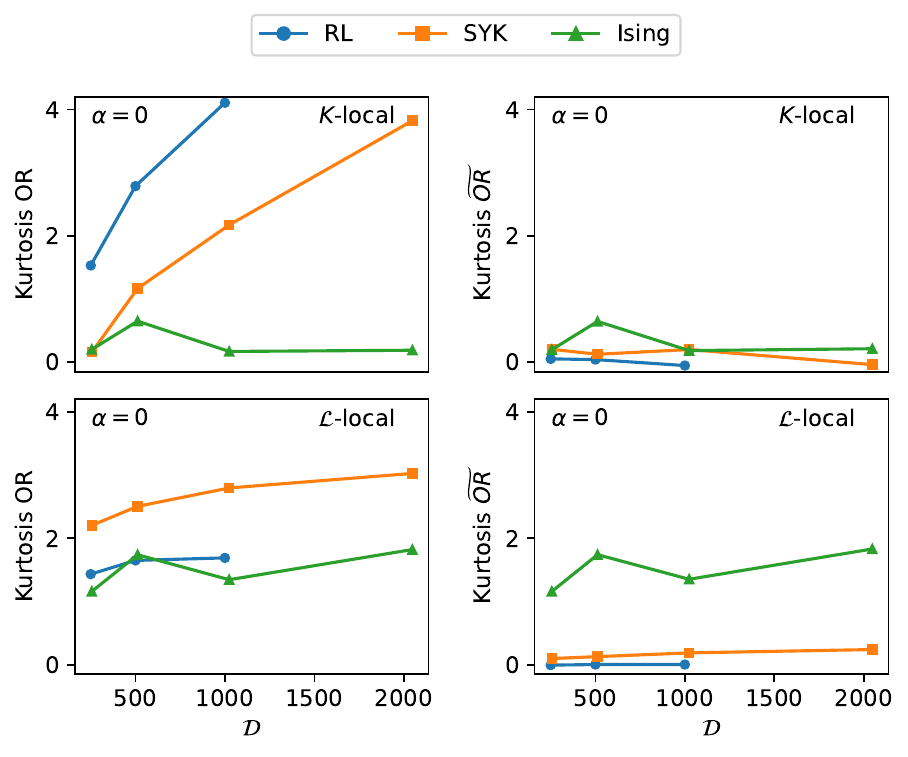}
    \caption{Excess kurtosis for the distributions in the first and third columns of Fig.~\ref{fig:alpha-0}. Only the quantity $\widetilde{OR}$ computed locally in the spectrum of $K$ shows small decreasing values of the kurtosis for all models.}
    \label{fig:alpha0-kurtosis}
\end{figure}

Furthermore, we analyze the finite-size effects in the distributions shown in Fig.~\ref{fig:alpha-0}. In Fig.~\ref{fig:alpha0-kurtosis}, we show the kurtosis for some of the distributions. Only $\widetilde{OR}$ computed locally in the spectrum of $K$ shows small decreasing values of the kurtosis for all models.

\end{document}